\documentclass[aps,prl,twocolumn,reprint,showpacs,superscriptaddress,nofootinbib]{revtex4}  % for double-spaced preprint
\usepackage{graphicx}  % needed for figures
\usepackage{mwe}
\usepackage{float}
\graphicspath{ {/Users/marta/Desktop/articolo/} }
\usepackage{dcolumn}   % needed for some tables
\usepackage{bm}        % for math
\usepackage{amssymb}   % for math
\usepackage{setspace}
\usepackage{blindtext}
\usepackage{enumitem}
\usepackage{hyperref}
\usepackage{textcomp}
\usepackage{nicefrac}
\usepackage{natbib}
\usepackage{amsmath}
\usepackage{amsfonts}
\begin{document}

\title{Bianchi I cosmology in the presence of a causally regularized viscous fluid
	}
	
\author{Giovanni Montani}
\email{giovanni.montani@frascati.enea.it}
\affiliation{ENEA, FSN-FUSPHY-TSM, R.C. Frascati, Via E. Fermi, 45 (00044) Frascati (RM), Italy.}
\affiliation{Dipartimento di Fisica, Universit\`a degli studi di Roma "La Sapienza", P.le A. Moro 5 (00185) Roma, Italy.}

\author{Marta Venanzi}
\email{m.venanzi@soton.ac.uk}
\affiliation{Dipartimento di Fisica, Universit\`a degli studi di Roma "La Sapienza", P.le A. Moro 5 (00185) Roma, Italy.}
\affiliation{University of Southampton, Department of Physics and Astronomy, Southampton, SO17 1BJ, United Kingdom.}

\date{\today}

\begin{abstract}
We analyze the dynamics of a Bianchi I cosmology in the presence of a viscous fluid, causally regularized according to the Lichnerowicz approach. We show how the effect induced by shear viscosity is still able to produce a matter creation phenomenon, meaning that also in the regularized theory we address, the Universe is emerging from a singularity with a vanishing energy density value. We discuss the structure of the singularity in the 
isotropic limit, when bulk viscosity is the only 
retained contribution. We see that, as far as viscosity is not a dominant 
effect, the dynamics of the isotropic Universe possesses 
the usual inviscid power-law behavior but in correspondence of an
effective equation of state, depending on the bulk viscosity 
coefficient. 
Finally, we show that, in the limit of a strong non-thermodynamical equilibrium of the Universe
mimicked by a dominant contribution of the effective 
viscous pressure, a power-law inflation behavior of the 
Universe appears, the cosmological horizons are removed and a significant amount of entropy is produced.
\end{abstract}

\pacs{04.20.Dw, 04.40.Nr}
\maketitle
\section*{Introduction}
The initial cosmological singularity has been demonstrated to be a true, generic property of the Universe \cite{BKL70,BKL82, Montani}. However, while the dynamics of the early Universe has been essentially understood, its physical and thermodynamical nature is far to be under control.
On one hand, quantum gravity effects are able of altering the standard dynamical features proposed in \cite{BKL82,K93,Mo95}, giving rise to fascinating alternatives (see for instance \cite{Ash},\cite{Pal}) and particle creation effects can also be relevant \cite{Star,Mo2001}. Calling for attention is also the string cosmology paradigm, as discussed for instance in \cite{Barrow}.
On the other hand, such an extreme region of evolution exhibits a rapid expansion and non-trivial out-of-equilibrium phenomena become possibly important, including the appearance of viscous features in the cosmological fluid. More specifically, one usually distinguishes the bulk viscosity from the shear viscosity: while the former accounts for the non-equilibrium effects associated to volume changes, the latter is a result of the friction between adjacent layers of the fluid. As a matter of fact, shear viscosity does not contribute in isotropic cosmologies, whereas it may significantly modify the dynamics of an anisotropic Universe, as we shall see below.
The simplest representation of a relativistic viscous fluid is provided by the so-called Eckart energy momentum tensor \cite{Eckart}. However, this formulation results to be affected by non-causal features, allowing the propagation of superluminar signals \cite{HL}. In order to amend such a non-physical behavior, a revised approach has been proposed by Israel \cite{Israel}, solving the non-causality problem via the introduction of phenomenological relaxation times. 
Having in mind the basic role the Bianchi I cosmology has 
in understanding the generic behavior of an inhomogeneous Universe near the singularity, in \cite{BK_Bianchi} and \cite{BK_Israel} such a model is studied as sourced by a viscous fluid, in both the Eckart and in the Israel approach respectively. One of the most intriguing issue coming out from such a study must be undoubtedly identified in the possibility of a singularity, from which the Universe emerges with 
negligible energy density and then a process of matter creation takes place. 
\\
In the present analysis, we face the study of this aforesaid peculiar solution in terms of an alternative causal regularization of the Eckart energy-momentum tensor, proposed by Lichnerowicz in \cite{LICH1}. 
Such a revised formulation is based on the introduction of the so-called \emph{index of the fluid}, de facto a regulator scaling the four-velocity field, so defining a \emph{dynamical velocity} of the fluid. 
This approach has been tested on some real systems, receiving interesting confirmation to its viability \cite{Disc1,Disc2}.
However, being derived via a phenomenological approach, the 
Lichnerowicz energy momentum tensor must be completed by 
the specification of an ansatz linking the 
fluid index to the thermodynamical variables of 
the system, so closing the dynamical problem. 
In what follows, we implement the Lichnerowicz treatment 
to the viscous Bianchi I cosmology, by pursuing 
two different tasks. On one hand, we study the 
solution with matter creation, fixing the fluid index via the 
request of incompressibility. Results show that the Universe evolves trough an intrinsic shear-driven anisotropic solution, meaning that also in the Lichnerowicz scenario the solution with matter creation exists and, actually, such a phenomenon is 
enhanced, being therefore not related to non-physical effects of the 
Eckart formulation. On the other hand, we analyze the isotropic limit near the singularity, by reducing the three scale factors of the Bianchi I model to be equal. The latter is known in literature as the flat Robertson-Walker Universe, for which only bulk viscosity may be relevant due to the homogeneity and isotropy of the model, preventing shear among different layers. In this specific study we see that, as far as bulk viscosity is not dominant, the regularization provided by the index of the fluid preserves the same power-law
behavior of the inviscid isotropic Universe. In other words, the bulk viscosity coefficient enters trough an effective equation of state, ranging the same parameters domain of an ideal fluid (i.e. between dust and stiff matter). Finally, we show how, if the bulk viscosity becomes sufficiently dominant, it is possible to get an equation of state having 
an effective polytropic index less than $\nicefrac{2}{3}$, leading to a power-law inflation solution. The latter is characterized by a massive entropy creation and no longer causal separation exists across the Universe regions.  
\section*{Basic formalism}
The Lichnerowicz original stress-energy tensor describing relativistic viscous fluids stands as follows \cite{LICH1}
\begin{align}
\label{TmunuC}
\begin{split}
T_{\mu\nu}&=(\rho+p)u_{\mu}u_{\nu}+pg_{\mu\nu}-\left(\varsigma-\frac{2}{3}\eta\right)\pi_{\mu\nu}\nabla_{\alpha}C^{\alpha}\\ &-\eta\pi_{\mu}^{\alpha}\pi_{\nu}^{\beta}\left(\nabla_{\alpha}C_{\beta}+\nabla_{\beta}C_{\alpha}\right),
\end{split}
\end{align}
where $\rho$ is the energy density, $p$ is the pressure, $g_{\mu\nu}$ denotes the metric tensor with the signature $(-+++)$ and $u^{\mu}$ is the four-velocity properly normalized as
\begin{equation}
u_{\mu}u^{\mu}=-1\,.
\end{equation} 
The bulk and shear viscous contributions are represented by the $\zeta$ and $\eta$ coefficients, respectively. Here,
\begin{equation}
\pi_{\mu\nu}=g_{\mu\nu}+u_{\mu}u_{\nu}
\end{equation}
is the projection tensor. Furthermore, $C^\mu$ represents the so-called dynamical velocity which is related to $u^\mu$ by 
\begin{equation}
\label{C}
C^{\mu}=Fu^{\mu},
\end{equation}
$F$ being the index of the fluid.\\
A simple algebra shows that expression (\ref{TmunuC}) can be rearranged as  
\begin{align}
\label{Tmunu}
\begin{split}
T_{\mu\nu}&=(\rho+p')u_{\mu}u_{\nu}+p'g_{\mu\nu} \\
&-\eta\,F\,\left[\nabla_{\mu}u_{\nu}+\nabla_{\nu}u_{\mu}+u_{\mu}u^{\alpha}\nabla_{\alpha}u_{\nu}+u_{\nu}u^{\alpha}\nabla_{\alpha}u_{\mu}\right],
\end{split}
\end{align}
where $p'$ is the total pressure containing the standard thermodynamical contribution and the negative component due to  viscosity, i.e.
\begin{align}
p'\equiv p-\lambda\nabla_{\alpha}C^{\alpha}, && \lambda\equiv\zeta-\frac{2}{3}\eta.
\end{align}
%Cita Lich
 The introduction of $F$ was first due by Lichnerowicz in order to describe viscous processes in relativistic dynamics, attempting to avoid superluminal signals. We can think of $F$ as a contribution which eliminates the non-causality features of the Eckart's formulation  by regularazing the velocity  ($F$ will be therefore referred below as the \emph{regulator} of the theory). The cosmological consequences of the Lichnerowicz description in isotropic cosmologies have been examined in \cite{Disc2}. In \cite{Disc1}-\cite{Disc4}, the index of the fluid is parametrized as
\begin{equation}
\label{F}
F=\frac{p+\rho}{\mu}\, ,
\end{equation}
where $\mu$ is the rest mass-density which satifies the conservation law
\begin{equation}
\label{restmass}
\nabla_{\alpha}(\mu u^{\alpha})=0.
\end{equation}
The main advantages of the Lichnerowicz theory have been pointed out in \cite{Disc2}. 
A key-point consists in the fact that expression (\ref{Tmunu}) reduces to the traditional description provided by Eckart upon setting $F=1$.
Furthermore, one of the main assumptions of the well-posedness theorems requires that \cite{Disc3,Disc4}
\begin{equation}
\label{F1}
F>1.
\end{equation}
Lichnerowicz was lead to introduce a new formulation for viscosity first because of the study of incompressible perfect fluids for which the following relation holds\\
\begin{equation}
\label{incomp}
\nabla_{\mu}C^{\mu}=0.
\end{equation}
%As we shall see below, this specific case is relevant near the singularity and actually it naturally fixes the behavior of the index of the fluid.
The basic aim of the present paper is to investigate the dynamical role of the 
	regulator $F$ in two relevant asymptotic regimes. 
	On one hand, we wondered how the solution with matter creation, derived 
	in \cite{BK_Bianchi} for the Bianchi I solution, 
	depends on the causal regularization of 
	the theory (i.e. does this effect survive
	in the presence of the Lichnerowicz formulation?). On the other hand, we are interested
	to a cosmologically relevant regime, corresponding to the flat isotropic limit,
	for which only the bulk viscosity must be 
	retained (the shear viscosity being suppressed due to  
	the isotropy hypothesis). 
	Both these regimes are investigated via a power-law solution, which is able to 
	capture the dominant term behavior in 
	the asymptotic solution. Such a technique offers a satisfactory answer to the questions above and follows the original treatment in \cite{BK_Bianchi,BK_Israel}. Nevertheless, the cosmological relevance
	of the isotropic case leads us to numerically investigate the dynamics of the system, even far from the initial singularity. 
	A subtle point in the Lichnerowicz approach to the regularization of a viscous fluid
	is the determination of the regulator $F$ in terms of the thermodynamical
	parameters. The two relevant regimes here addressed require different descriptions
	for the regulator expression. The most delicate case is the one associated to the matter creation, for which both the energy density and the Universe volume vanish
	asymptotically to the singularity. 
	Then, the construction of a reliable ansatz for large asymptotic values of $F$ 
	appears a non-trivial task. Nonetheless, a simple solution to this puzzle is 
	offered by the condition of incompressibility defined by (\ref{incomp}), which, as we shall see below, is also appropriate for the comparison to the original analysis in \cite{BK_Bianchi}. 
	The isotropic flat Universe case can be instead easily faced by retaining the same choice as in \cite{Disc1}, i.e. $F$ is provided by the ratio between 
	the enthalpy and mass density of the Universe as stated in (\ref{F}). This formulation appears
	well-grounded owing to the fact that the thermal history of the Universe naturally 
	ensures that $F$ is large in the early Universe and it tends to unity in the 
	present stage of evolution (say for a redshift $z<100$, when the pressure term 
	becomes negligible).
\noindent
\section*{Field equations}
The line-element of the Bianchi I model in the synchronous reference frame reads as
\begin{equation} 
\label{metricaBI}
ds^{2}=-dt^{2}+R_1(t)^2\,dx^{2}+R_2(t)^2\,dy^{2}+R_3(t)^2\,dz^{2}
\end{equation}
and hence the metric determinant is given by the following relation
\begin{equation}
\sqrt{-g}=R_1 R_2 R_3\equiv R^{3}.
\end{equation}
Above, $x, y, z$ denote Euclidean coordinates and $R_1(t)$, $R_2(t)$ and $R_3(t)$ are dubbed cosmic scale factors. As well-known (see \cite{Montani},\cite{Landau1} and \cite{Gravitation}) such a model describes in vacuum an intrinsic anisotropic Universe, but in the presence of matter it can also admit the isotropic limit \cite{KirillovMo}. Let us now introduce the following quantity
\begin{align}
\label{H}
\begin{split}
H&\equiv\left(ln\,R\right)^{.}\\
&=\frac{1}{3}\left(\frac{\dot{R_1}}{R_1}+\frac{\dot{R_2}}{R_2}+\frac{\dot{R_3}}{R_3}\right).
\end{split}
\end{align}
where the dot denotes the time derivative. %1/3a+b+c
In order to have a compatible system, we set up the Einstein equations in a comoving frame with the matter source in which $u^{0}=1$ and $u^{i}=0$, $i=1,2,3$.\\
In this frame, $p'=p-\lambda\dot{F}-3 \lambda F H$ and the stress-energy tensor components are
\begin{align}
\label{Tmunu_comp}
\begin{split}
T^0_0&=-\rho, \\
T^i_i&=p'-2\eta F \frac{\dot{R_i}}{R_i}.
\end{split}
\end{align}
Thus, the Einstein equations for the Bianchi I spacetime, having the viscous Lichnerowicz tensor as source (here only the $ii$ and $00$-components are non-vanishing), can be written as
\begin{align}
\frac{\ddot{R_2}}{R_2}+\frac{\ddot{R_3}}{R_3}+\frac{\dot{R_2}\dot{R_3}}{R_2 R_3}&=-\chi\left(p'-2\eta F\frac{\dot{R_1}}{R_1}\right), \label{Einst1}  \\ 
\frac{\ddot{R_1}}{R_1}+\frac{\ddot{R_3}}{R_3}+\frac{\dot{R_1}\dot{R_3}}{R_1 R_3}&=-\chi\left(p'-2\eta F\frac{\dot{R_2}}{R_2}\right), \label{Einst2}  \\ 
\frac{\ddot{R_1}}{R_1}+\frac{\ddot{R_2}}{R_2}+\frac{\dot{R_1}\dot{R_2}}{R_1 R_2}&=-\chi\left(p'-2\eta F\frac{\dot{R_3}}{R_3}\right), \label{Einst3} \\ 
\frac{\dot{R_1}\dot{R_2}}{R_1 R_2}+\frac{\dot{R_1}\dot{R_3}}{R_1 R_3}+\frac{\dot{R_2}\dot{R_3}}{R_2 R_3}&=\chi\rho, \label{Einst4} 
\end{align}
$\chi$ being the Einstein constant.\\
From the spatial components (\ref{Einst1},\ref{Einst2},\ref{Einst3}), it is possible to show that the system admits the following integrals of motion 
\begin{equation}
\label{integrale1}
\frac{\dot{R_{i}}}{R_{i}}=H+s_{i}R^{-3}e^{\varphi}.
\end{equation}
Here $\dot{\psi}\equiv -2\chi\eta F$ and the quantities $s_i$ are such that $s_{1}+s_{2}+s_{3}=0$.
The evolution of $H$ is obtained using the trace of the $ii$-components combined with the $00$-component (\ref{Einst4}) of the Einstein equation so getting
\begin{equation}
\label{H_punto}
\dot{H}=\chi\rho-\frac{1}{2}\chi h-3H^{2}+\frac{3}{2}\chi\zeta FH+\chi(\frac{1}{2}\zeta-\frac{1}{3}\eta)\dot{F},
\end{equation}
where 
\begin{equation}
\label{enthalpy}
h=\rho+p
\end{equation}
represents the specific enthalpy.
Moreover, the hydrodynamic equations $\nabla_{\nu}T_{\mu}^{\nu}=0$ (only the $0$-component is not vanishing here) provide the following evolution for $\rho$ 
\begin{equation}
\label{rho_punto}
  \dot{\rho}=-4\chi\eta F\rho-3Hh+9H^{2}\varsigma F+12H^{2}\eta F+3H\varsigma\dot{F}-2H\eta\dot{F}.
\end{equation} 
The first integrals (\ref{integrale1}) can be re-cast in a more compact form by the use of (\ref{H_punto}) and (\ref{Einst4}) as 
\begin{equation}
\label{integrale2}
\rho=\frac{1}{\chi}\left(3H^{2}-q^{2}R^{-6}e^{2\varphi}\right),
\end{equation}
where $q^{2}\equiv \frac{1}{2}\left(s_{1}^{2}+s_{2}^{2}+s_{3}^{2}\right)$.
It worth noting that setting $q^2=0$ corresponds to the isotropic case.
Equations (\ref{H}), (\ref{H_punto}), (\ref{rho_punto}) and (\ref{integrale2}) together with the $ii$-components of the field equations represent the full set of the dynamical equations characterizing the present model. In order to close the system we introduce a polytropic index $\gamma$ and we consider an equation of state of the form
\begin{align}
h=&\gamma\rho, && 1\leq\gamma\leq2
\end{align}
where, e.g., $\gamma=1$ corresponds to dust matter and $\gamma=\frac{4}{3}$ to the radiation cases, respectively.
\noindent
\section*{Asymptotic solutions with matter creation}
\raggedbottom
As well known \cite{BKL70}, approaching the initial singularity, the Bianchi I solution in vacuum is Kasner-like and the presence of a perfect fluid is negligible. In \cite{BK_Bianchi}, it has been shown instead that in the presence of a shear viscous contribution the situation significantly changes but a Kasner-like solution still exists and it is characterized by a vanishing behavior of the energy density.
Here, we want to verify if such a peculiarity survives when the Eckart representation of the viscous fluid is upgraded in terms of the Lichnerowicz causal reformulation. It is rather easy to realize via a simple asymptotic analysis, that the ansatz in \cite{Disc1} fails in the region of small value of $\rho$. In fact, the latter predicts $F\sim\rho R^3$ and clearly vanishes near the singularity if $\rho\sim0$ (as $R^3$ is going naturally to zero toward the singularity). In investigating the solutions in this region we need therefore to search for a different representation of $F$. A possibility is to infer its form by picking the case of an incompressible fluid.
From equation (\ref{incomp}) and by using the line-element (\ref{metricaBI}), one immediately gets the following expression for $F$ 
\begin{equation}
F=\frac{F_{0}}{R^{3}},
\end{equation}
where we used the convention for which the subscript number denotes an integration constant. The latter is a convention which we will use throughout this paper.
Clearly, $F$ grows to infinity as approaching the singularity, without contradicting the constraint (\ref{F1}) and assuring well-behaving solutions without superluminal signals. In this regard, it is rather natural to realize that the value of $F$ should significantly grow in extremely relativistic regimes as near the cosmological singularities. As suggested in \cite{BK_Bianchi}, we assume that in the region of low density the viscous coefficients can be expressed as power-laws of the energy density with exponents greater than unity, i.e.,
\begin{align}
\eta&=\eta_{1}\rho^{\alpha_{1}}, && \zeta=\zeta_{1}\rho^{\beta_{1}}, &&\rho\rightarrow 0, && \alpha_{1}\geq1, \beta_{1}\geq1. \label{visc1}
\end{align}
It is worth noting that for an incompressible fluid the $\zeta$-terms automatically drop out from the field equations. This is also the case treated in \cite{BK_Bianchi} because there the bulk viscosity contributions is asymptotically negligible for small energy densities, taking $\beta_{1}>\alpha_{1}$. 
In our analysis, in order to search for a consistent solution, we assume that the density vanishes faster than the volume $R^{3}$ as the system approaches the singular point $(H,\rho)=(+\infty,0)$ in the $(H,\rho)$ plane\footnote{Details about the dynamical study of the asymptotic solutions in the Eckart representation can be found in \cite{BK_Bianchi}.}. 
Moreover, in the right-handed side of equation (\ref{H_punto}) we can retain  the quadratic term in $H$ only. Then, the limiting form of the equations (\ref{H_punto}) and (\ref{rho_punto}) are
\begin{align}
\dot{H}&=-3H^{2} \label{Hpunto2}, \\ 
\dot{\rho}&=-3\gamma\rho H+18\eta_{0}F_{0}\frac{\rho^{\alpha_{1}}}{R^{3}}H^{2} \label{rhopunto2}.
\end{align}
It is easy to see that $\dot{\varphi}\rightarrow0$ as it contains a positive power-law of the energy density and hence we can fix, without loss of generality,  $\varphi\equiv 0$. The constraint equation (\ref{integrale2}) reduces to the following relation 
\begin{equation}
\label{integrale3}
H=\sqrt{3}qR^{-3}.
\end{equation}
Then, one can easily check that the leading order asymptotic solutions for $t\rightarrow0$ of the simplified Einstein equations (\ref{Hpunto2}), (\ref{rhopunto2}), (\ref{integrale3}) read as
\begin{align}
H&=\frac{1}{3t}, && \rho=Kt^{\frac{2}{\alpha_{1}-1}}, && R^{3}&=\sqrt{3}qt,\\
R_{i}&=\left(\sqrt{3}qt\right)^{p_{i}}, && p_{i}=\frac{1}{3}+\frac{s_{i}}{3q}
\end{align}
where  $K$ is a constant depending on the parameters of the model. In order the solutions to be asymptotically self-consistent, the relations above are applicable only when $\alpha_{1}>3$, slightly different with respect to the results given by the Eckart approach in \cite{BK_Bianchi} where $\alpha_{1}>1$. 
It is worth noting that we found that the energy density in this causal model decays more rapidly to zero with respect to the non-regularized case studied in \cite{BK_Bianchi} by Belinskii and Khalatnikov (BK) where
\begin{equation}
\rho^{BK}\sim t^{\frac{1}{\alpha_{1}-1}},\qquad\alpha_{1}>1.
\end{equation}
In other words, we see how the Lichnerowicz approach leads to a cosmological model in which the universe emerges from the singularity with a greater matter-rate creation with respect  to the Eckart case.
\section*{Viscous dynamics in the isotropic Universe}
We now focus our investigation in the opposite regime where the energy density takes a diverging value near the singularity.
Then, let us consider the isotropic limit of the solution, leading to the flat Robertson-Walker Universe, by setting $q^2=0$ in the first integral (\ref{integrale2}). The latter reduces to
\begin{equation}
\label{integrale3}
\rho=\frac{1}{\chi}3H^{2}.
\end{equation}
As a natural consequence of isotropy and compatibility issues with the Einstein equations, shear viscosity is not permitted in the model and bulk viscosity is the only retained contribution in the energy momentum expression (\ref{Tmunu}).
Indeed, in a homogeneous Universe, isotropically expanding, no friction between different layers can occur and the shear viscosity can affect inhomogeneous perturbations only.
This set up has already been examined in \cite{BK_FRW} for an Eckart fluid via the dynamical system approach, and  in \cite{Disc1} via the Lichnerowicz treatment in an attempt to explain dark-energy related current issues. Here we drew our attention to the behavior of the solutions in the limit of large density values.
We address the problem by parameterizing $F$ according to expression (\ref{F}) or,
\begin{equation}
F=\frac{\gamma \rho R^3}{\mu_0},
\end{equation} 
where we have used the fact that $\mu=\mu_0 {R}^{-3}$ because of the rest-mass conservation law (\ref{restmass}). It is easy to check that the time derivative of $F$ yields
\begin{equation}
\dot{F}=\left(3H+\frac{\dot{\rho}}{\rho}\right)F.
\end{equation}
Then, in this limiting case, the Einstein equations take the form
\begin{align}
\dot{H}&=\chi\left(1-\frac{1}{2}\gamma\right)\rho-3H^{2}+3\chi\zeta  FH+\frac{1}{2}\chi\zeta F\frac{\dot{\rho}}{\rho}, \label{Hpunto3}  \\ 
\dot{\rho}&=-3H\gamma\rho+18H^{2}\varsigma F+3H\varsigma  F\frac{\dot{\rho}}{\rho}. \label{rhopunto3}
\end{align}
For the bulk viscosity coefficient in the limit of large density, we still have a power-law behavior of the type:
 \begin{align}
 \zeta=\zeta_{2}\rho^{\beta_{2}}, && \rho\rightarrow\infty, && 0\leq\beta_{2}\leq\frac{1}{2}\,. \label{visc2}
 \end{align}
Notice that we are using a different subscript with respect to equation (\ref{visc1}), corresponding to a different asymptotic region of the energy density. 
 The solutions of the full-set of the field equations given by (\ref{H}), (\ref{integrale3}), (\ref{Hpunto3}) and (\ref{rhopunto3}) are investigated assuming that the energy density and the volume are evolving like powers of time
 according to the relations 
 \begin{align}
 \label{timepowers}
 \rho=\rho_{0}t^{y}, && R^{3}=R_{0}^{3}t^{x}, && y<0, && x>0,
 \end{align}
 as the time $t\rightarrow 0$. As a consequence, the viscosity evolves trough the following expression
 \begin{equation}
 \zeta=\zeta_2{\rho_0}^{\beta_{2}}{t}^{\beta_{2} y}.
 \end{equation}
 From equation (\ref{timepowers}) with the use of (\ref{H}) we immediately get
 \begin{equation}
 \label{Htimepower}
 H=\frac{x}{3t}.
 \end{equation}
Using equation (\ref{Hpunto3}) and being $\dot{H}\sim t^{-2}$, it is necessary that $y=-2$ in order the system to be self-consistent. Similar arguments lead us to conclude that
 \begin{equation}
 \label{x_beta}
  x=2\beta_{2}+1.
 \end{equation}
 Then, one finds the following evolutions in terms of $t$ for the energy density and the scale-factor
 \begin{align}
 \rho=\frac{x^{2}}{3\chi}t^{-2}, \label{rho4}\\
 R={R_0} t^{\frac{2}{3\gamma_{eff}}}, && \gamma_{eff}\equiv\frac{2}{2\beta_{2}+1}. \label{gammaeff}
 \end{align}
 where we introduced an effective equation of state $\gamma_{eff}$. One can check that $\gamma$ is related to $\beta_{2}$ via the following expression
 \begin{equation}
 \label{gamma1}
\gamma=\frac{2}{2\beta_{2}+1}\left[\frac{1}{1-4\beta_{2}(2\beta_{2}+1)^{\beta_{2}}\frac{\zeta_{2}R_{0}^{3}}{(3\chi)^{\beta_{2}}\mu_{0}}}\right].
 \end{equation}
 Furthermore, one can see that the regulator evolves with time as 
 \begin{equation}
 \label{F_t}
 F\sim t^{2\beta_{2}-1}
 \end{equation}
which is asymptotically growing when $0\leq\beta_{2}<1/2$ and reduces to a positive constant when $\beta_{2}=1/2$, leading to well-behaving regularized solutions. An interesting additional feature stands out from the behavior of the scale factor $R$. When the Robertson-Walker geometry is coupled to an ideal fluid the volume typically evolves as $R_{RW}\sim t^{\frac{2}{3\gamma}}$. Here we observe that we still have an isotropic limit, as it is found in \cite{BK_Bianchi}, but instead of being negligible, the viscosity acquires a fundamental role in the dynamics of the Universe, driving the evolution trough an effective equation of state.
 Indeed, the range of the possible values of $R$ in (\ref{gammaeff}) perfectly coincides with the standard non-viscous homogeneous and isotropic universe. In particular, 
 \begin{itemize}[noitemsep]
 \item for $\beta_{2}=1/2$ we have $R\sim t^{\nicefrac{2}{3}}$: the universe maps a dust-dominated Friedmann Universe with an effective equation of state $\gamma_{eff}=1$;
 \item for $\beta_{2}=0$ we get $R\sim t^{\nicefrac{1}{3}}$: the solution evolves toward a stiff-matter dominated Universe with an effective equation of state $\gamma_{eff}=2$;
 \end{itemize}
 which allows us to conclude that the introduction of $F$ in the Friedmann universe has the role to encode all the viscous effects in the dynamics and what we see at the end is the usual ideal fluid equation of state. We say in this sense that bulk viscosity is regularized.\\ 
 Now we emphasize that by extrapolating our solutions to the regime $\beta_{2}>\nicefrac{1}{2}$ we obtain an intriguing dynamical property of the Universe. In fact, from equation (\ref{gammaeff}) for $\beta_{2}=1/2+\varepsilon$ with $\varepsilon>0$ we immediately get
 \begin{equation}
  \gamma_{eff}=\frac{1}{1+\varepsilon}\rightarrow p=-\frac{\varepsilon}{(1+\varepsilon)}\rho.
 \end{equation}
 For $\varepsilon>\nicefrac{1}{2}$ (i.e. $\beta_{2}>1$ and $\gamma_{eff}<\nicefrac{2}{3}$) this dynamical behavior corresponds to a powerlaw inflation solution, induced by a negative effective pressure $p<-\nicefrac{1}{3}\rho$. Indeed, it is easy to realize that the cosmological horizon 
 \begin{equation}
d_h(t)\equiv R(t) \int_{0}^{t}\frac{dt'}{R(t')}
 \end{equation} 
takes in this case a divergent value. In this scenario the Universe corresponds to a unique causal region and we can think of it as a viable solution to the horizon paradox.
 It is worth noting that, since we are dealing with 
 an isotropic flat Universe, we get 
 $H\sim \sqrt{\rho}$ (see equation (\ref{integrale3})). Then, the restriction 
  $\beta_{2}< \nicefrac{1}{2}$ in the expression (\ref{visc2}), derived for an Eckart representation of the fluid ($F\equiv 1$), acquires a clear physical meaning. 
 In fact, as far as such a restriction holds, the negative effective pressure, 
 due to bulk viscosity, behaves like 
 $\rho^{\beta_{2}+\nicefrac{1}{2}}<\rho$, for large $\rho$ values. 
 This means that the standard (positive) 
 thermodynamical pressure $p = (\gamma - 1)\rho$ 
 remains always the dominant contribution. 
 This is coherent with the idea that the
 bulk viscosity representation of non-equilibrium 
 effects is valid only on a perturbative level. Clearly, the presence of the regulator $F$ slightly changes this situation since it enhances the weight of viscosity in the dynamics and this is at the ground of the present results. Nonetheless, the regime $\beta_{2}>1$ can be 
 qualitative interpreted as a fluid-like representation 
 for strong non-equilibrium effects, not surprising 
 in the limit of the singularity, when the 
 geometrical velocity of Universe collapse diverges. Thus, the power-law inflation solution we find in such an extreme regime can be thought as 
 the qualitative feature induced by a 
 cosmological continuous source, 
 whose thermodynamical evolution can not be 
 approximated via equilibrium stages. 
 If we accept a fluid representation for such a 
 limiting scenario, we can argue that the 
 superluminar geometrical velocity of the early 
 phases of the Universe expansion is able 
 to open the horizon size, making the 
 cosmological space causally connected as a whole.\\
 It is now worthwhile to stress a remarkable feature of the obtained solution which makes it essentially different from the same limit treated in \cite{BK_Bianchi}. Indeed, there, the isotropic solution was considered for $t\rightarrow 0_-$ simply because it corresponds to a singular point in the dynamical system approach of equations (\ref{H}), (\ref{H_punto}), (\ref{rho_punto}) and (\ref{integrale2}).
 Here we deal with increasing $t$ values and our dynamics describes an expanding universe from the initial singularity. This choice characterizing the evolution regime is physically allowed when one considers the behavior of the entropy per comoving volume. The latter has the form
 \begin{equation}
 \sigma \sim \rho^{\frac{1}{\gamma}}R^{3}\sim t^{2 \left(\frac{1}{\gamma_{eff}}-\frac{1}{\gamma}\right)}.
 \end{equation} 
 Since from equation (\ref{gamma1}) we see that
  \begin{equation}
  \label{gamma2}
  \gamma=\gamma_{eff} \left[\frac{1}{1-4\beta_{2}(2\beta_{2}+1)^{\beta_{2}}\frac{\zeta_{1}R_{0}^{3}}{(3\chi)^{\beta_{2}}\mu_{0}}}\right]>\gamma_{eff},
  \end{equation}
  it is straightforward to infer that the entropy per comoving volume increases due to dissipation processes when the universe expands. This brings us to conclude that we are dealing with a cosmological paradigm in which the universe emerges from the singularity causally connected as a whole and with a significant entropy creation (this effect is enhanced by increasing $\beta_{2}$ values). This suggests how extreme non-equilibrium thermodynamics near the singularity could play a relevant role in solving some unpleasant paradoxes of the standard cosmological model, namely the horizon and entropy ones.
  Further remarks should now be done about the obtained solutions (\ref{Htimepower}), (\ref{rho4}), (\ref{gammaeff}) and (\ref{F_t}). Despite the power-law approach given by (\ref{timepowers}) is able to provide a satisfactory characterization of the asymptotic behavior of the model under consideration, the underlying cosmological relevance which comes with it leads us to support the solutions with an additional numerical investigation. 
  	In this regard, we show in the right panel of figure (1) how the power-law approximation is largely predictive near enough to the singularity, while discrepancies take over as the Universe volume expands. On the other hand, one expects that far from the singularity we assist to a gradual decrease of the role of the viscosity. In figure (1) (left panel) it is shown that the numerical solution for large times tends to overlap the standard inviscid flat Robertson-Walker dynamics. The cosmological picture coming out is consistent with the paradigm of an Universe characterized by a viscous non-equilibrium dynamics close to the Big-Bang whose viscous features are suppressed as the volume expands, recovering the isentropic homogeneous and isotropic Universe  (as a consequence of the
  	decreasing value of the energy density).
  \begin{figure*}[t]
  	\centering	
\begin{tabular}{p{0.5\textwidth} p{0.5\textwidth}}
	\vspace{0pt} \includegraphics[width=0.47\textwidth, height=7.5cm]{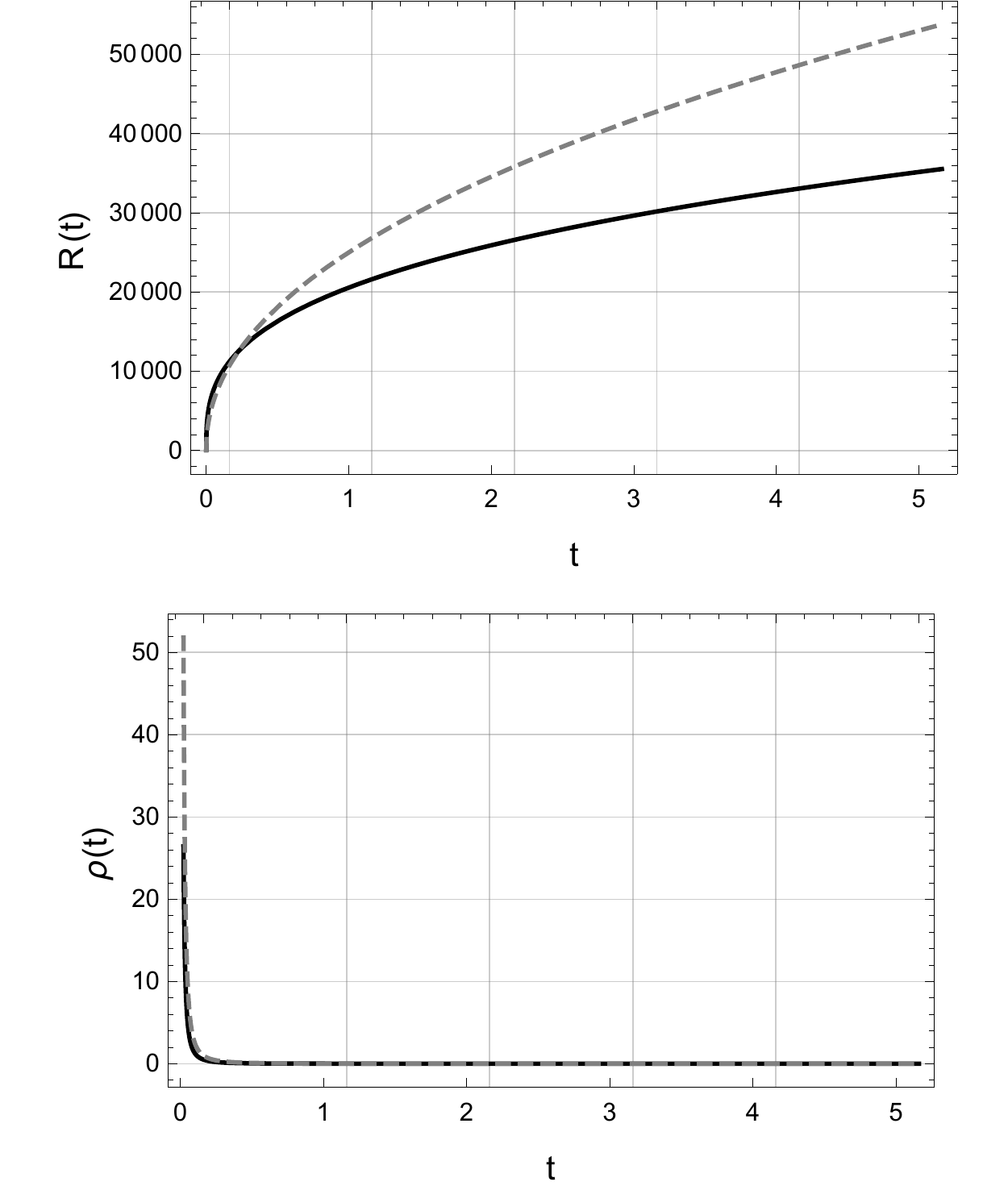} &
	\vspace{0pt} \includegraphics[width=0.47\textwidth,height=7.5cm]{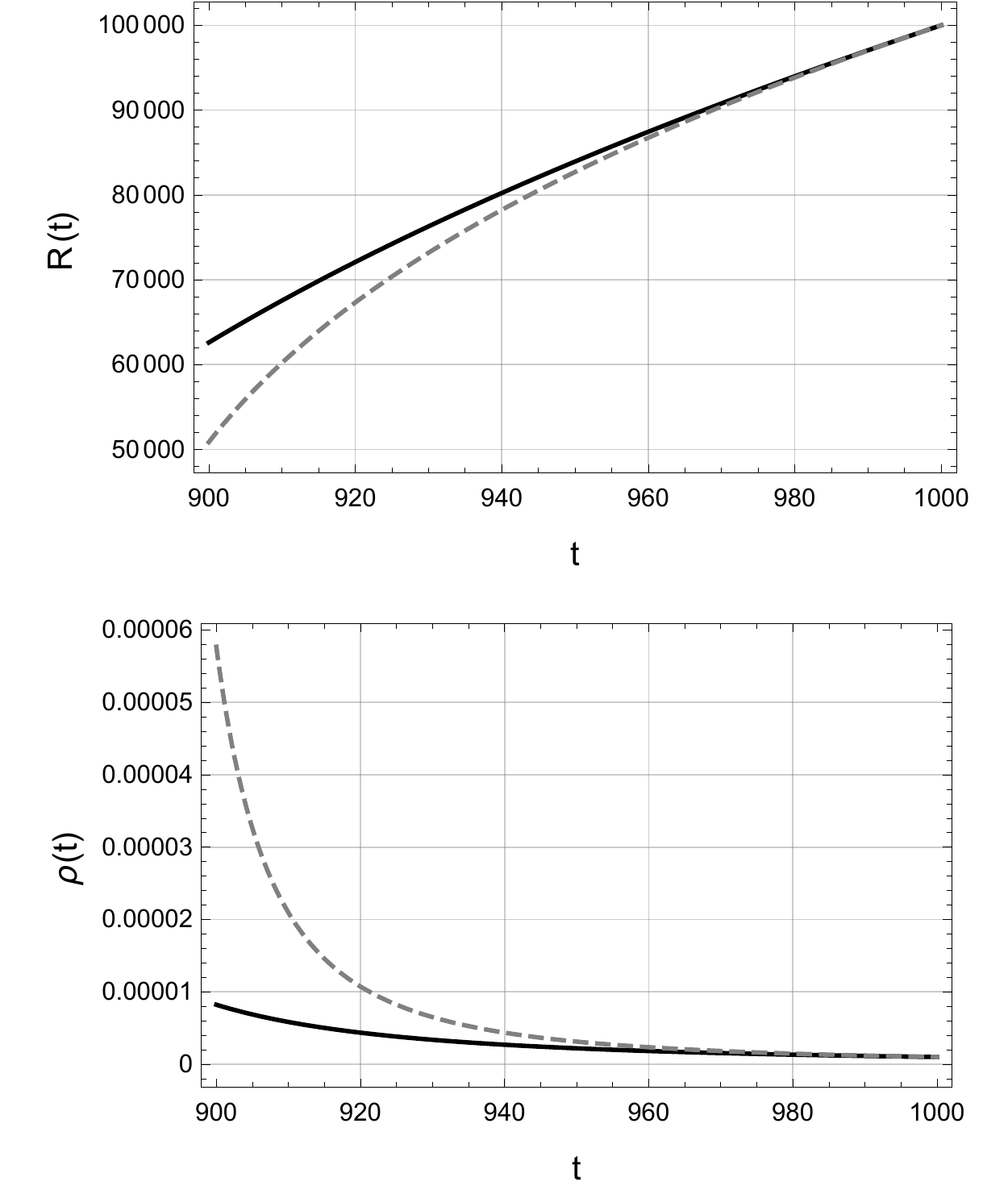}
\end{tabular}
\caption{(left) Numerical integration of the viscous isotropic model (continuous line). The power-law approximation, given by the dashed line, is shown to give reliable predictions of the dynamics of the Universe near the singularity. (right) Far from the singularity evolution of the viscous isotropic model. We see how the effect of viscosity is suppressed with time and the solutions gradually approaches the inviscid Friedmann Universe (dashed lines).}
\end{figure*}
  The numerical analysis performed above enforces the idea that strong viscous effects (happening for $\gamma _{eff}<2/3$) can be viewed as a consistent solution to the horizon and entropy paradox, by means of a power-law dynamics in the very early Universe evolution.
We further make some considerations about the stability of the obtained solutions. In this regard, we observe that in \cite{carlmont} the problem of the stability of a flat isotropic Universe has been analyzed  
 	in presence of bulk viscosity, according to 
 	the Eckart formulation. 
 	In the direction forward in time 
 	(the same one we are interested here), it 
 	 is shown how the Universe is stable under 
 	scalar perturbations. This result is 
 	expected to be maintained in the present formulation too, simply because the
 	regulator $F(t)$ is large near the singularity 
 	(according to the ansatz (\ref{F}). 
 	By other words, one can infer that the 
 	presence of $F$ simply enhances the effect 
 	fo the viscosity, preserving the stability 
 	properties derived in \cite{carlmont}. 
 	This conjecture acquires a reliable meaning in the considered power-law approximation of the solution, where the presence 
 	of the regulator can be easily restated in terms of an effective value for
 	$\beta_{2}$ in the corresponding Eckart formulation (i.e. asymptotically to the
 	singularity $F$ becomes a power-law 
 	in the energy density of the viscous fluid).
 	We conclude this section by observing 
 	that the power-law solution (\ref{Htimepower}), (\ref{rho4}), (\ref{gammaeff})  
 	can be extended to the negative and 
 	positive curved Robertson-Walker 
 	geometry \footnote{We remind the reader that the isotropic Bianchi I model actually 
 	is the flat Robertson-Walker 
 	Universe}, as far as the effective polytropic index $\gamma_{eff}$ remains
 	greater   
 	than $2/3$. Under such a restriction, 
 	the spatial curvature term (behaving like 
 	$R^{-2}$), is asymptotically negligible 
 	near the singularity with respect to both
 	the energy density $\rho$ and the viscous 
 	contribution. 
 	The situation is different in the case 
 	of the power-law inflation, when 
 	$\gamma_{eff} < 2/3$, since the curvature term 
 	can play a relevant role. However, this is 
 	a typical feature of the inflationary-like solutions, whose existence requires the 
 	spatial gradients to be sufficiently smooth 
 	\cite{KT90}.
 	From a physical point of view, if we 
 	cut our asymptotic regime at the Planck 
 	time (assuming that before the quantum 
 	dynamics is concerned), we could 
 	require that the spatial curvature is, at that time, sufficiently 
 	small and then it will become negligible 
 	forward in time.
 \section*{Conclusions}
 We have studied the influence of a viscous cosmological fluid on the Bianchi I Universe dynamics in the neighborhood of the initial singularity. The characterizing aspect of the presented analysis consists of describing the viscous fluid via the Lichnerowicz formulation, introducing a causal regulator 
 (the index of the fluid) in order to ensure a causal dynamics. We have also pointed out how this additional new degree of freedom must be properly linked to the geometrical and thermodynamical variables. We have investigated the role of the two viscosity coefficients in two different limits, when only one of them is dynamically relevant. 
 The shear viscosity has been taken into account in the case of an incompressible fluid, 
 evaluating the modification that the regulator introduces in the well-known
 solution with matter creation derived in 
 \cite{BK_Bianchi}. We have shown that such a peculiar phenomenon, characterized by an asymptotically vanishing energy density, not only survives in the Lichnerowicz
 formulation but it is actually enhanced. The presence of a bulk viscosity 
 term has been analyzed in the isotropic limit of the Bianchi I cosmology and again asymptotically to the initial singularity. We derived a power-law
 solution, which outlines some interesting 
 features: i) the standard non-viscous Friedmannian behavior is encountered when bulk viscosity is a small deviation from equilibrium ($\beta_{2}<\nicefrac{1}{2}$), but this time the fluid presents an equation of state with an effective dependence on viscosity; 
 ii) when bulk viscosity fully dominates the dynamics with allowance made for strong non-equilibrium effects ($\beta_{2}>1$), we see that the universe evolves trough a power-law inflation solution to the initial singularity, implying the divergence of the cosmological horizon and the subsequent disappearance 
 of the Universe light-cone. 
 Entropy production has been addressed in the isotropic Robertson-Walker limit and specifically for dominant viscosity, where entropy tremendously grows.
 Despite one may argue whether or not the above-mentioned fluid description is possible in this extreme regime of dominant bulk viscosity, the issues above strongly suggest that a comprehensive understanding of the Universe birth and of the so-called horizon and entropy paradox can 
 not be achieved before a clear account of 
 the non-equilibrium thermodynamical evolution 
 near the singularity will be properly provided.  
 \section*{Acknowledgements}
 This paper has been developed within the CGW collaboration (www.cgwcollaboration.it) and supported by the \textit{TornoSubito} project.  We would like to thank Riccardo Moriconi for his valuable assistance in the numerical integration of the model.
 M. V. is thankful to Dr. Shabnam Beheshti who significantly motivated and assisted this work and to Queen Mary, University of London for providing all the necessary facilities. 
  
\end{document}